\documentclass[10pt,journal,compsoc]{IEEEtran}
\ifCLASSOPTIONcompsoc
  \usepackage[nocompress]{cite}
\else
  \usepackage{cite}
\fi
\ifCLASSINFOpdf
\usepackage[pdftex]{graphicx}
\else
\fi
\usepackage{amsmath}
\usepackage{amsmath}
\usepackage{amsfonts}
\usepackage{amssymb}
\usepackage{amsthm}

\usepackage{hyperref}

\usepackage{algpseudocode}

\ifCLASSOPTIONcompsoc
  \usepackage[caption=false,font=footnotesize,labelfont=sf,textfont=sf]{subfig}
\else
  \usepackage[caption=false,font=footnotesize]{subfig}
\fi

\usepackage{booktabs}

\usepackage{xcolor} 

\begin{document}
%
\title{Enabling Homomorphically Encrypted Inference for Large DNN Models}

\author{Guillermo~Lloret-Talavera,
        Marc~Jorda,
        Harald~Servat,
        Fabian~Boemer,
        Chetan~Chauhan,
        Shigeki~Tomishima,
        Nilesh~N.~Shah,
        and~Antonio~J.~Pe\~na%
\IEEEcompsocitemizethanks{\IEEEcompsocthanksitem G. Lloret-Talavera,
  M. Jorda, and A. J. Pe\~na are with the Barcelona Supercomputing
  Center (BSC).\protect\\
E-mail: \{guillermo.lloret,marc.jorda,antonio.pena\}@bsc.es
\IEEEcompsocthanksitem H. Servat, F. Boemer, C. Chauhan,
  S. Tomishima, and N. N. Shah are with Intel Corporation.\protect\\
E-mail:
\{harald.servat,fabian.boemer,chetan.chauhan,shigeki.tomishima, nilesh.n.shah\}@intel.com\protect\\
\copyright~2021 IEEE.  Personal use of this material is permitted.  Permission from IEEE must be obtained for all other uses, in any current or future media, including reprinting/republishing this material for advertising or promotional purposes, creating new collective works, for resale or redistribution to servers or lists, or reuse of any copyrighted component of this work in other works.}}

%
%

\markboth{Manuscript accepted for publication in IEEE Transactions on Computers}%
{Lloret-Talavera \MakeLowercase{\textit{et al.}}: Enabling Homomorphically Encrypted Inference for Large DNN Models}
\IEEEtitleabstractindextext{%
\begin{abstract}
The proliferation of machine learning services in the last few years
has raised data privacy concerns.
Homomorphic encryption (HE) enables inference using encrypted data but it incurs  100x--10,000x memory and runtime overheads.
Secure deep neural network (DNN) inference using HE is currently limited by computing and memory resources, with frameworks requiring hundreds of gigabytes of DRAM to evaluate small models.
To overcome these limitations, in this paper we explore the
feasibility of leveraging hybrid memory systems comprised of DRAM and
persistent memory.
In particular, we explore the recently-released
Intel\textsuperscript{\textregistered}~Optane\textsuperscript{\texttrademark}
PMem technology and the
Intel\textsuperscript{\textregistered}~HE-Transformer
nGraph\textsuperscript{\textregistered} to run large neural networks
such as MobileNetV2 (in its largest variant) and ResNet-50 for the
first time in the literature.
We present an in-depth analysis of the efficiency of the executions with different hardware and software configurations.
Our results conclude that DNN inference using HE incurs on friendly
access patterns for this memory configuration, yielding efficient
executions.
\end{abstract}

}

\maketitle

\IEEEdisplaynontitleabstractindextext

%
\IEEEpeerreviewmaketitle

\IEEEraisesectionheading{\section{Introduction}\label{sec:introduction}}

%
%
%
%
\IEEEPARstart{M}{achine} learning (ML) enables solving problems that are infeasible with traditional techniques in fields like computer vision or speech recognition.
Although the available computational power is ever-increasing,
most ML projects are still relatively highly compute-intensive.
Data scientists and commercial ML deployments often resort to
third-party providers to speed up training and inference tasks. For instance,
mobile-based voice recognition is frequently implemented as a cloud service.


Practical homomorphic encryption (HE) implementations emerged
recently and enable computations directly on encrypted data; the
(encrypted) result is correct as if it were produced by the
traditional method (decryption, computation, and encryption) and may
be decrypted using the secret key. HE is not exempt from
drawbacks that render it currently impractical in many scenarios: the
size of the data increases fiercely when encrypted, whereas the
computation time is considerably higher than that over unencrypted
data. HE requires the selection of several parameters; ensuring high
security level, such as post-quantum, requires large memory and
runtime overheads~\cite{chase2017security}. Values for this overhead
vary depending on the security parameters, usually being in the order
of 100x for compute and 100x--10,000x for data size.
There are clear use cases for deep neural network (DNN) model and
dataset encryption, such as intellectual property or sensitive data
protection. While customary RAM, based on DRAM technology, is far from
able to entirely host production-sized homomorphically encrypted DNNs
or associated datasets, out--of--band algorithms will suffer from
intrinsic data movement overheads, apart from the added code
complexity. Only reduced-size homomorphically encrypted DNN inference cases have been reported, on top of DRAM, while training is considered too time-consuming~\cite{boemer2019ngraph-he}.

Recently, dual in-line memory modules (DIMMs) based on new technologies have become commercially available.
One of these alternatives to traditional DRAM is the Intel\textsuperscript{\textregistered}~Optane\textsuperscript{\texttrademark} PMem product line~\cite{OPTANE_DC}.
Besides non-volatility, it offers a much larger capacity than DRAM, with up to 512~GB of memory in a single DIMM.
However, the access latency for the persistent memory is considerably higher than
that of DRAM, especially when storing data.
With respect to DRAM, its latencies increase 2x--6x for reads and 6x--30x for writes depending on the
access pattern, whereas bandwidth decreases around 75\% for reads and
90\% for writes.
To palliate this
large gap with DRAM performance, which is exacerbated at non-sequential access patterns due to large access
block sizes, DRAM and PMem are usually combined on a hybrid memory system.
This technology may be an enabler for large DNNs to be run using HE in
a single machine; however, prior to this work, it was unclear how Intel Optane PMem latency characteristics would affect performance.

In this work, we present, for the first time in the literature, an
exhaustive performance analysis of an HE framework running on a system
with hybrid DRAM + persistent memory subsystems.
We present the results of our experiments including different DRAM
capacities, but also the largest DNNs reported
to date on an HE inference framework (namely MobileNetV2 on its
largest variant and ResNet-50), as enabled by the use of persistent memory technology.
We have analyzed the performance of Intel Optane PMem to determine
its viability for this specific use case. Our results reveal that HE
inference yields a friendly access pattern to Intel's implementation
of persistent memory technology, hence enabling for the first time the execution of
large DNN models leveraging HE.

In summary, the contributions of this article are: (1) We report the
largest DNN models run to date using HE; (2) We report for the first
time the use of persistent memory technology to enable large DNN inference
leveraging HE; and (3) We provide the corresponding novel in-depth
analysis of the viability of using persistent memory technologies for this specific
use case.

The rest of the document is structured as follows.
Section~\ref{sec:background} provides the necessary background.
Section~\ref{sec:related} discusses related work in the
literature. Section~\ref{sec:setup} describes the testbed used in our
experiments. Section~\ref{sec:results} presents our results and their
analysis. Section~\ref{sec:conclusions}
reviews the conclusions of this work.

\section{Background}
\label{sec:background}

This section is intended to provide readers with the necessary
background information to understand the rest of the manuscript.
First we introduce HE. Next, we discuss the main
features of the persistent memory implementation we leveraged in this study.

\subsection{Homomorphic Encryption}

HE is a type of encryption that enables limited computation on the ciphertext, without use of the secret key.
This feature allows data to remain confidential when being processed in an untrusted environment.

HE was proposed in 1978~\cite{rivest1978data}.
Over the next 30 years, researchers discovered a variety of partial HE
(PHE)
schemes---supporting a single operation, such as addition or
multiplication; somewhat HE (SHE) schemes---supporting several operations, such as both addition and multiplication, but on only a subset of circuits; and leveled HE (LHE) schemes---supporting arbitrary circuits, up to a limited size or depth.
In our work, we focus on the CKKS LHE scheme~\cite{cheon2017homomorphic}.

The security of many HE schemes, including CKKS, is based on the ring learning with error (RLWE) problem~\cite{lyubashevsky2010ideal}, whose security derives from the hardness of the shortest vector problem (SVP) in lattices.
The RLWE problem uses polynomials in $\mathcal{R}_q := \mathbb{Z}_q[X] / (X^N +1)$, where the \emph{polynomial modulus} $N$ is typically a power of two and the \emph{coefficient modulus} $q$ is a prime number.
Concretely, $\mathcal{R}_q$ contains polynomials of degree $N-1$ whose coefficients are integers modulo $q$.
Addition and multiplication in $\mathcal{R}_q$ may be performed via regular polynomial addition and multiplication, followed by reduction by $X^N+1$, and coefficient-wise reduction modulo $q$.
Given polynomials $a_i(x), s(x)$ drawn uniformly at random from $\mathcal{R}_q$, and $e_i(x)$, drawn from a small error distribution, typically a discrete Gaussian, the RLWE problem is to determine the secret polynomial $s(x)$ given several samples $(a_i, a_i \cdot s + e_i)$.
The choice of $N$ and $q$ determine the security level $\lambda$ of the RLWE problem.
A security level of $\lambda$ bits indicates ${\sim} 2^\lambda$ operations are required to break the decryption, with typical values of $\lambda \in \{128, 192, 256\}$.

The presence of the noisy polynomials $e_i(x)$ yields noisy HE operations.
Furthermore, the noise grows as additional HE operations are performed.
Once the noise has reached a certain threshold, the homomorphism between encrypted and plaintext operations breaks down and decryption yields an inaccurate result.
Noise growth is a fundamental difficulty in scaling use of HE in practice.

In 2009, Gentry discovered a bootstrapping procedure, which removes noise from the ciphertext.
By applying bootstrapping to an LHE scheme, Gentry constructed the first fully homomorphic encryption (FHE) scheme, supporting an unlimited number of additions and multiplications~\cite{gentry2009fully}.
Using polynomial approximations, it became possible to compute arbitrary functions on encrypted data.
While initial implementations of bootstrapping could last several minutes for a single ciphertext, recent HE schemes~\cite{ducas2015fhew,chillotti2017faster} operating on Boolean circuits, rather than arithmetic circuits, enable bootstrapping on the order of milliseconds.
However, performing addition and multiplication in Boolean circuits requires several Boolean operations, creating additional overhead.

LHE schemes such as CKKS often employ the residue number system (RNS) representation to efficiently perform arithmetic on large integers.
Using RNS representation, the coefficient modulus $q$ is factored into several primes, $q = \prod_{i=0}^{L-1} q_i$ for some depth $L$.
For convenience, we denote $q = \{ \lfloor \log_2(q_0) \rfloor, \lfloor \log_2(q_1) \rfloor, \hdots, \lfloor \log_2(q_{L-1}) \rfloor \}$, that is, $q$ as a list of bit-widths of each coefficient modulus.
CKKS also uses approximate arithmetic, such that $Dec(c_1 \cdot c_2 + c_3) \approx m_1 \cdot m_2 + m_3$ for ciphertexts $c_i=  Enc(m_i)$.
As a result, the RNS form of the CKKS scheme~\cite{cheon2018full} features superior arithmetic circuit performance compared to the BFV scheme~\cite{brakerski2012fully,fan2012somewhat}.

\subsection{Intel\textsuperscript{\textregistered}~Optane\textsuperscript{\texttrademark} PMem}

Recently, in Q2 2019, Intel released a non-volatile, byte-addressable
memory in the form of DIMMs, named
Intel\textsuperscript{\textregistered}~Optane\textsuperscript{\texttrademark}
Persistent Memory or
Intel\textsuperscript{\textregistered}~Optane\textsuperscript{\texttrademark}
PMem\footnote{Formerly known as
  Intel\textsuperscript{\textregistered}~Optane\textsuperscript{\texttrademark} DC Persistent Memory.}, which is compatible with 2\textsuperscript{nd} Generation Intel\textsuperscript{\textregistered} Xeon\textsuperscript{\textregistered} Scalable processors.
This type of memory sits between memory and storage, delivering the best of two worlds through the convergence of memory and storage product traits.
The persistent memory modules may co-exist with traditional DDR4 DRAM DIMMs on the same platform, and run up to 2,666~MT/s while operating at 12, 15, or 18~W.
As of time of this writing, Intel offers persistent modules in sizes from 128 GB to 512 GB, which allows 2-socket platforms to store up to 6 TB of data.

Systems equipping Optane PMem feature two operating modes: Memory Mode
and App Direct Mode.
Memory Mode aims at exposing a system with a huge volatile memory capacity while being completely
transparent to applications and operating systems, and hence the persistence attribute is lost.
In contrast, in App Direct mode, the software sees the DRAM and the persistent memory as two distinct memory pools and may choose what data to place into each tier.

In Memory Mode, DRAM is managed by the CPU memory controller as a direct-mapped write-back cache.
Consequently, and as expected, data locality plays a relevant role in terms of
performance, because access latency is the same as DRAM when an access
hits, but it has to pay for the cost of accessing DRAM and persistent memory when the access misses in DRAM.
The access to the Intel Optane PMem is optimized for 256~byte transfers and since the CPU requests data in chunks of 64~bytes, the DIMM controller integrates a media prefetch buffer of the recently accessed 256~bytes to rapidly respond to the CPU if data is already found in this buffer.

Compatible CPUs integrate new performance counters for the analysis of Intel Optane PMem-enabled software at different architectural levels~\cite{CLX_perfcntrs}.
The core counters are able to identify which data addresses have been
referenced and which part of the memory system (including PMem) provided the data.
The uncore counters provide information such as the traffic volume observed by the CPU memory controllers on the different memory types.
The SMART counters may be used to determine the locality of the software media access patterns.

\section{Related Work}
\label{sec:related}

Privacy-preserving machine learning (PPML) has gained interest in recent years.
Given a machine learning model $M$, and input data $x$, typically owned by separate parties, one goal of PPML is to perform inference, i.e. compute $M(x)$, while reducing the data leakage among parties. Previous work typically uses one or more of several privacy-preserving
primitives, including secure multi-party computation (MPC), HE, differential privacy, and more recently, functional secret sharing~\cite{ryffel2020ariann}. Differential privacy seeks to minimize the leakage of $M$ or $x$ resulting from observing the result $M(x)$. In contrast, cryptographic techniques such as MPC and HE seek to minimize the data leakage \emph{during} the computation process.

MPC-based DNN inference includes SecureML~\cite{mohassel2017secureml}, XONN~\cite{riazi2019xonn}, Secure\-NN~\cite{wagh2018securenn}, ABY3~\cite{mohassel2018aby3}, PySyft~\cite{ryffel2018generic}, TF-Encrypted~\cite{dahl2018private}, and CrypTFlow~\cite{kumar2019cryptflow}.
HE-based DNN inference was introduced by the seminal CryptoNets paper~\cite{gilad2016cryptonets}, which performed inference on a 5-layer network on the MNIST dataset.
Subsequent work has improved the performance, scaled to larger networks, and integrated with DNN compiler technologies~\cite{boemer2019ngraph-he,boemer2019ngraph-he2,hesamifard2017cryptodl,boura2018chimera,dathathri2019chet}.
While, due to DRAM limitations, the largest model reported to date running inference using HE is
MobileNetV2 (a lightweight network with a small number of parameters)
reduced by an expansion factor of 0.35 from the original size~\cite{boemer2019ngraph-he2}, we report results running MobileNetV2 on its
largest expansion factor plus, for the first time in the literature, a
fully-featured DNN: ResNet-50.
Several works also combine multiple privacy-preserving primitives, such as HE with MPC~\cite{juvekar2018gazelle,boemer2020mp2ml}.

One difficulty in using HE for DNN inference is the choice of plaintext packing, which enables encoding of multiple scalars into a single plaintext.
HE operations on the plaintext, as well as the encrypted ciphertext, apply to each scalar, in a single-instruction multiple data (SIMD) manner.
Given a tensor of shape $N \times C \times H \times W$, \emph{batch-axis packing} encodes the tensor as a $C \times H \times W$-ciphertext tensor, each storing $N$ scalars.
\emph{Inter-axis packing} uses a different encoding scheme, for instance as an $N$-ciphertext tensor, each storing $C \times H \times W$ scalars.
In general, batch-axis packing attains high throughput at the cost of high latency and high total memory requirement, whereas inter-axis packing attains low latency and memory usage at the cost of low throughput.
Previous work has used both batch-axis packing~\cite{boemer2019ngraph-he,gilad2016cryptonets,boemer2019ngraph-he2} and inter-axis packing~\cite{juvekar2018gazelle,jin2019carenets,pmlr-v97-brutzkus19a}.
The choice of batch-axis packing requires hundreds of gigabytes of memory for MobileNetV2~\cite{boemer2019ngraph-he2}, preventing scaling to larger networks on conventional DRAM systems.

A number of recent software efforts to develop and
optimize HE primitives on top of different architectures exist~\cite{sealcrypto,dai2015cuhe,chou2018faster,roy2019fpga};
however, only a few very recent DNN frameworks are
known supporting HE~\cite{boemer2019ngraph-he,boemer2019ngraph-he2,dathathri2019chet,van2019sealion} and these are tuned for
DRAM-friendly access patterns. To date, no previous report exists of DNN
inference leveraging HE on top of non-DRAM memory technology.

\section{Experimental Setup}
\label{sec:setup}

In this section, we describe the technical details of our experimental
setup, including our testbed system, the inference engine we used, and
the neural network models that served as our use cases.

\subsection{Testbed}

\begin{figure}\centering
    \subfloat[DRAM-only, 192 GB DRAM (DO)]{\includegraphics[width=0.9\linewidth]{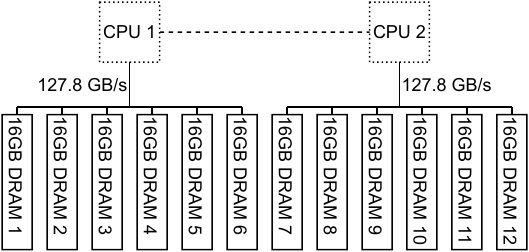}%
    \label{fig:conf_do}}
  \\
	\subfloat[Memory Mode, 96 GB DRAM + 6 TB Intel Optane (MM96)]{\includegraphics[width=0.9\linewidth]{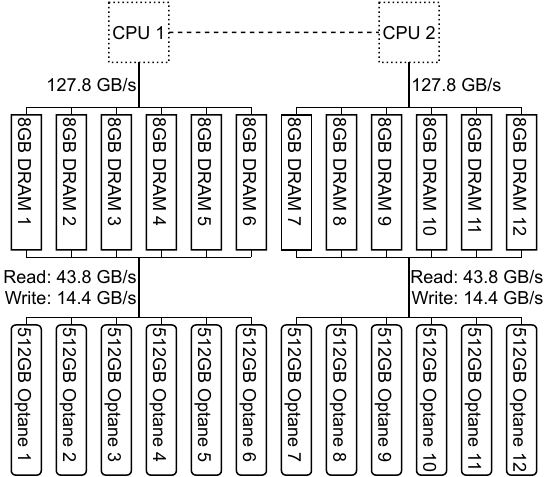}%
    \label{fig:conf_mm96}}
  \\
    \subfloat[Memory Mode, 32 GB DRAM + 6 TB Intel Optane (MM32)]{\includegraphics[width=0.9\linewidth]{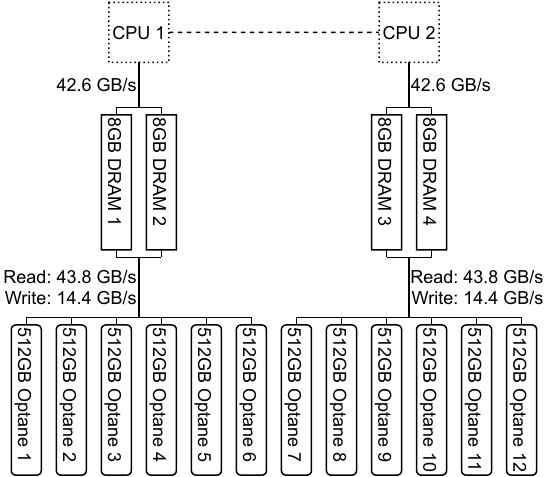}%
    \label{fig:conf_mm32}}
  \caption{Logical view of the DO and MM memory configurations.}\label{fig:confs}
\end{figure}

Our experiments are performed in a dual socket system using
Intel\textsuperscript{\textregistered}~Xeon\textsuperscript{\textregistered}
Platinum 8260L processors, with 24 cores on each socket running at a nominal frequency of 2.30~GHz.
For all experiments involving persistent memory, twelve 512~GB Intel Optane PMem DIMMs were used.
Each of these modules features a theoretical bandwidth of 7.3~GB/s for
read accesses and 2.4~GB/s for write accesses~\cite{OPTANE_BW}.
For the Memory Mode (MM) experiments, we have worked with two different configurations.
The first configuration (MM32) uses four 8~GB DRAM DIMMs, while the
second (MM96) uses twelve DIMMs.
For the DRAM-only experiments (DO), we have used twelve 16~GB DRAM DIMMs.
Figure \ref{fig:confs} shows an overview of the MM32, MM96 and DO configurations.
Both DRAM DIMM models use DDR4 and feature a theoretical bandwidth of 21.3~GB/s.

Equipping reduced DRAM space and bandwidth on the MM32
configuration, we demonstrate the viability of our target use case on
an energy-friendly memory configuration, being the energy consumption
per byte of DRAM about 10 times higher than that of the Optane PMem (375
mW/GB vs. 35~mW/GB).

Our system runs Fedora 27 (Workstation Edition) with kernel version 4.18.8.
All the software used was compiled from sources using GCC 7.3.1 targeting the native architecture.

\subsection{Inference Engine}
\label{sec:hetransformer}

Intel\textsuperscript{\textregistered}~HE-Transformer for nGraph\textsuperscript{\texttrademark}~\cite{boemer2019ngraph-he2} is an HE backend to the Intel nGraph Compiler~\cite{cyphers2018intel}, a graph compiler and runtime for artificial neural networks.
This backend supports the CKKS encryption scheme and relies on the simple encrypted arithmetic library (SEAL)~\cite{sealcrypto} for the implementation.
Operations that involve comparison (ReLU, MaxPool, etc.) are not supported natively in the CKKS scheme.
To perform these operations, HE-Transformer uses a client-aided protocol in which the server sends the encrypted data to the client, which decrypts the data, performs the operation, encrypts the output, and sends it back to the server.
This approach acts as a bootstrapping process to refresh the ciphertext noise, while also performing the comparison operation.
However, it may leak the model weights to the client.
This leakage may be mitigated using garbled
circuits~\cite{boemer2020mp2ml} but this approach currently
suffers from an even higher runtime overhead, so we do not consider it
in this work.
HE-transformer enables data scientists to train networks on the
hardware of their choice, then easily perform inference on encrypted
data using popular deep learning frameworks such as TensorFlow.

In our experiments we have used HE-Transformer v0.6.0, which features
the following dependencies: TensorFlow v1.14.0, nGraph v0.25.0,
nGraph-bridge v.0.18.0, and SEAL v3.3.1.

We use the same encryption parameters as~\cite{boemer2019ngraph-he2} in all of our experiments, namely:
\begin{itemize}
  \item Polynomial modulus degree: $N=4096$
  \item Security level: $\lambda=128$
  \item Coefficient modulus: $q=\{30, 22, 22, 30\}$
\end{itemize}

As in~\cite{boemer2019ngraph-he2}, we consider the use case of a plaintext model and encrypted data.

\subsection{Neural Network Models}
We make use of two well-known neural network models as our use cases:
MobileNetV2 and ResNet-50.

\subsubsection{MobileNetV2}

MobileNetV2~\cite{s2018mobilenetv2} is a neural network model that uses depth-wise separable convolution to reduce the model size and complexity.
Thanks to these layers, MobileNetV2 requires roughly 9 times less computation than comparable neural networks, making it ideal for use in mobile and embedded systems.
This efficiency makes it particularly suitable for HE, since the main
disadvantage of this type of encryption is its high overhead.
The MobileNetV2 architecture features two modifiable parameters:

\begin{itemize}
  \item Width Multiplier: The number of channels in each layer with respect to the original model.
  \item Input Resolution: Size of width and height of the square image received by the network.
\end{itemize}
We denote a specific MobileNetV2 architecture with a tuple: (width multiplier, input resolution).
The higher the parameter values, the greater the accuracy obtained but at the cost of larger memory footprint and longer computation time.
Different configurations of both parameters have been tested in our experiments.

\subsubsection{ResNet-50}

Residual Network (ResNet) is a popular family of convolutional neural networks used for many computer vision applications~\cite{he2016deep}.
These networks are easier to optimize than traditional DNNs, which translates into shorter training times.
ResNet-50 is 50 layers deep and may classify images with a resolution
of $224\times 224$ pixels into 1,000 different categories.
In the ImageNet dataset~\cite{imagenet_cvpr09}, it attains a top 1
accuracy of 75\% and a top 5 accuracy of 92\%.

\subsection{Profiling Tools}

We have leveraged two profiling frameworks to confirm whether the application is using the PMem efficiently:
Extrae~\cite{EXTRAE}\slash Paraver~\cite{PARAVER}, and
Intel\textsuperscript{\textregistered}~VTune\textsuperscript{\texttrademark} Platform Profiler~\cite{PLATFORM_PROFILER}.
Extrae is a library that monitors parallel applications (using OpenMP,
MPI, and pthread, among others) and emits an activity record on a Paraver trace-file.
Paraver is a post-mortem visualization tool for qualitative and quantitative analysis.
Intel VTune Platform Profiler, on the other hand, is both the collector and visualizer for a system-wide profiling tool that allows obtaining a holistic view of system behavior including CPU, memory, network, and disk usage.
While both tools collect PMem-related metrics through the CPU performance counters, they have some fundamental differences.
Extrae/Paraver are targeted to explore a single-application
performance by means of instrumentation and sampling, which enables a
precise characterization, leading the resulting trace-file size to depend on the application activity. 
Intel VTune Platform Profiler samples system activity periodically, which allows monitoring longer runs independently of the application activity, but the tool does not attribute performance to specific routines.

\section{Experimental Analysis}
\label{sec:results}
In this section, we present and analyze the results of our in-depth
performance evaluation. We first focus our attention on the
MobileNetV2 model. Next, we discuss the ResNet-50 network.

\subsection{MobileNetV2}
\label{sec:mobilenetresults}

We have used the pre-trained MobileNetV2 models offered by TensorFlow\footnote{\url{https://github.com/tensorflow/models/tree/master/research/slim/nets/mobilenet}}, which vary in the width multiplier (0.35, 0.5, 0.75, 1.0, 1.3, and 1.4) and input resolution (96, 128, 160, 192, and 224).
For the two largest width multiplier settings only the 224 input resolution is available.

\begin{figure*}[ht]\centering
  \begin{minipage}{.49\textwidth}\centering
    \includegraphics[width=1\textwidth]{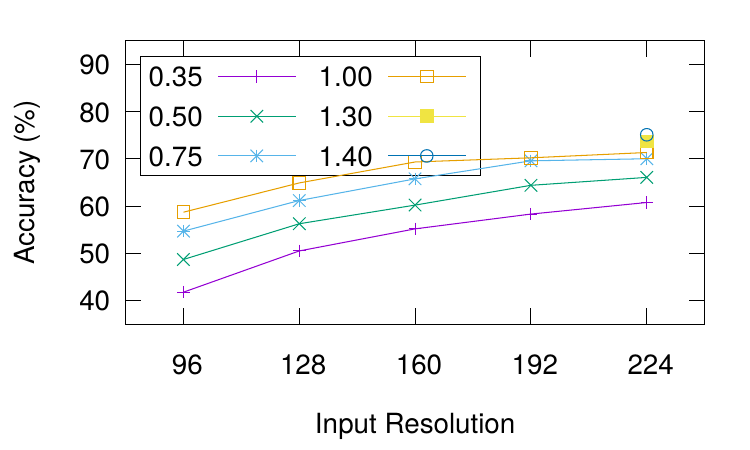}
    \caption{Top 1 accuracy.}\label{fig:top1}
  \end{minipage}
  \begin{minipage}{.49\textwidth}\centering
    \includegraphics[width=1\textwidth]{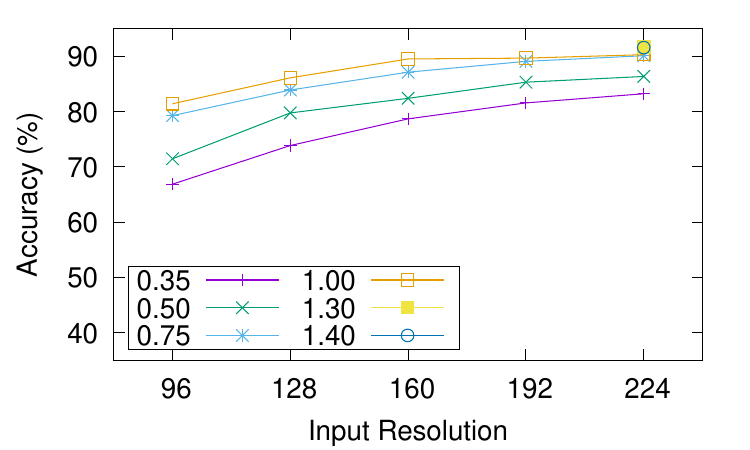}
    \caption{Top 5 accuracy.}\label{fig:top5}
  \end{minipage}
  \begin{minipage}{.49\textwidth}\centering
    \includegraphics[width=1\textwidth]{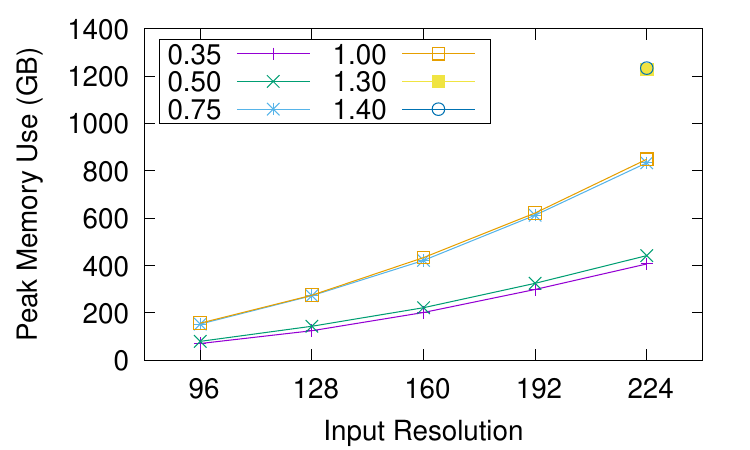}
    \caption{Memory footprint.}\label{fig:mem}
  \end{minipage}
  \begin{minipage}{.49\textwidth}\centering
    \includegraphics[width=1\textwidth]{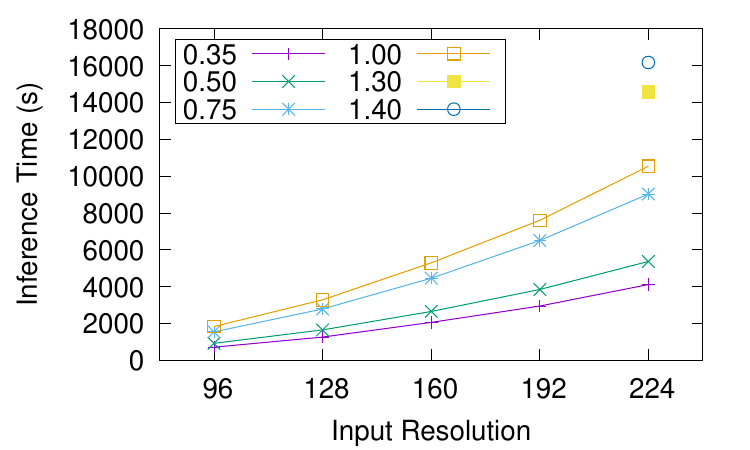}
    \caption{Inference time (MM32).}\label{fig:32time}
  \end{minipage}
\end{figure*}

We performed experiments with all possible configurations to determine
the memory consumption and the accuracy (top 1 and top 5) of each
network when leveraging HE.
We chose a batch size of 2,048 because it is the maximum possible value for the chosen encryption parameters.
Figure~\ref{fig:top1}, Figure~\ref{fig:top5}, and Figure~\ref{fig:mem}, show the results of these executions.
As expected, both the memory footprint and the obtained accuracy increase with larger models.
The largest model attains a top 1 accuracy of 75.1\% and top 5 accuracy
of 91.6\%, requiring 1.2~TB of memory.
Large models have not been feasible to be run in discrete systems so far
because of these memory requirements.
Thanks to recently-emerged persistent memory technology, now we have the amount of
memory required for this task; however, the feasibility of running inference on top
of the added latency this technology poses remained to be seen.

We have obtained the run times as an average of ten repetitions.
The maximum relative standard deviation is 3.2\%.
Figure~\ref{fig:32time} shows the times with the MM32
configuration.
The DO configuration has been used as a baseline,
although only the six smallest models fit within 192~GB of RAM,
starting at 71~GB.
Table~\ref{tab:cmp_mm_dram} shows the execution times of these models.
The DO configuration (using 192 GB DRAM) is merely up to 
4\% faster than MM96 (using 96 GB DRAM as cache) and
11\% faster than
MM32 (using 32 GB DRAM as cache),
despite the considerable raw performance difference of the main memory subsystem leveraged by the latter with respect to DRAM. 
The MM32 configuration populates only 2 of the 6 available DRAM DIMMs, yielding a third of the bandwidth compared to the DO configuration, which is fully populated.
In a memory--bandwidth--bound application, we would expect the execution times to
be further impacted.
However, the executions with the fully populated system are only about 10\%
faster, which indicates that memory bandwidth is not the main
bottleneck, being a positive first indicator of the feasibility of
leveraging persistent memory technologies as an enabler for large DNN inference with HE.

The additional latency incurred by the memory controller to
manage the cache in MM is also part of this time difference.
We have evaluated the latency incurred by the memory controller when in MM through
Intel\textsuperscript{\textregistered}~MLC\footnote{\url{https://software.intel.com/content/www/us/en/develop/articles/intelr-memory-latency-checker.html}})
and observed that local and remote socket accesses experience 10\% and 6\% overhead, respectively.

\begin{table}
  \caption{MobileNetV2 time comparison (MM vs. DO)}
  \centering
  \begin{tabular}{crrrr}
      \toprule
    \multicolumn{1}{c}{\begin{tabular}[c]{@{}c@{}}Model
        \end{tabular}} &
    \multicolumn{1}{c}{\begin{tabular}[c]{@{}c@{}}Memory\\
        Usage (GB)\end{tabular}} &
    \multicolumn{1}{c}{\begin{tabular}[c]{@{}c@{}}Time (s)\\ DO
        \end{tabular}} &
    \multicolumn{1}{c}{\begin{tabular}[c]{@{}c@{}}Time (s)\\ MM96
        \end{tabular}} &
    \multicolumn{1}{c}{\begin{tabular}[c]{@{}c@{}}Time (s)\\ MM32
        \end{tabular}} \\
    \midrule
    (0.35,  96) &  71 &   650 &   675 &   714 \\
    (0.35, 128) & 125 & 1,158 & 1,183 & 1,262 \\
    (0.50,  96) &  80 &   842 &   867 &   923 \\
    (0.50, 128) & 144 & 1,503 & 1,537 & 1,653 \\
    (0.75,  96) & 153 & 1,393 & 1,452 & 1,545 \\
    (1.00,  96) & 157 & 1,621 & 1,691 & 1,833 \\
    \bottomrule
    \end{tabular}
  \label{tab:cmp_mm_dram}
  \end{table}

\subsubsection{Analysis with Paraver}
\label{subsec:Paraver}

Since HE-Transformer exploits parallelism through
OpenMP\cite{dagum1998openmp} to deploy the most time-consuming parts
of the library, we benefit from the Extrae abilities to automatically instrument this runtime to monitor the application behavior.
We have additionally
instrumented manually a number of functions of interest, such as: convolution,
multiplication, reshape, etc.
We correlate with Paraver the instrumented functions in HE-Transformer with several performance metrics, including CPU hardware counters.
We have collected the following performance counters to analyze the performance of the Intel Optane PMem in Memory Mode:

\begin{itemize}
  \item MEM\_LOAD\_RETIRED.LOCAL\_PMM: Retired load instructions with local persistent memory as the data source and the data request missing L3 and DRAM cache.
  \item MEM\_LOAD\_L3\_MISS\_RETIRED.REMOTE\_PMM: Retired load instructions with remote persistent memory as the data source and the data request missing L3 and DRAM cache.
  \item MEM\_LOAD\_L3\_MISS\_RETIRED.LOCAL\_DRAM: Retired load instructions
    whose data sources missed L3 but were serviced from local DRAM.
  \item MEM\_LOAD\_L3\_MISS\_RETIRED.REMOTE\_DRAM: Retired load
    instructions whose data sources missed L3 but were serviced from remote DRAM.
\end{itemize}

Despite using sampling mechanisms,
the long running times of the largest models generate
overwhelmingly large traces.
For this reason, only three small models have been profiled in-depth
for MobileNetV2: (0.35, 96), (0.35, 128), and (0.75, 96).

Table~\ref{tab:32MM_075_96} shows the result of the largest MobileNetV2 model
profiled (0.75, 96) on the MM32 configuration.
Since the values for the local and remote sockets were balanced, these are grouped in the table with a value that aggregates the number of accesses across the sockets.
We have only explored the operations of the inference phase for the time measurement.
As we will discuss in Section~\ref{subsec:PlatformProfiler}, there is an initialization phase at the beginning of the network execution in which HE-transformer executes a number of optimization passes.
Consequently, we have decided to exclude this phase because in a production
scenario it would be executed once, while the inference phase would be invoked multiple times.

The main difference between the smallest and largest models is the
absolute number of accesses to DRAM and PMem spaces.
The percentage of DRAM accesses for Convolution decreases slightly (46\%) in the smallest model while the percentage for the BoundedRelu increases (49\%).
The ratios of PMem accesses, the main focus of our study, are very similar to the smallest models using the same DRAM configuration.

\begin{table*}
  \caption{Extrae metrics running MobileNetV2 (0.75, 96) on the MM32
    configuration}

  \centering
  \begin{tabular}{l|rr|rr|rr|r}
    \multicolumn{1}{c|}{Function} & \multicolumn{2}{c|}{Time (s)} & \multicolumn{2}{c|}{DRAM (K-Loads)} & \multicolumn{2}{c|}{PMem (K-Loads)} & \multicolumn{1}{c}{DRAM/PMem} \\ \hline
    \toprule
    Add               & 26,605           & 0.78\%    & 6,333                 & 2.23\%       & 1,345              & 6.02\%         & 4.75             \\
    AvgPool           & 1,077            & 0.03\%    & 80                   & 0.03\%       & 10                & 0.05\%         & 7.65             \\
    BoundedRelu       & 125,401          & 3.70\%    & 88,402                & 30.90\%      & 17,198             & 76.92\%        & 5.14             \\
    Concat            & 10,578           & 0.31\%    & 115                  & 0.04\%       & 5                 & 0.03\%         & 20.29            \\
    Constant          & 24,898           & 0.73\%    & 664                  & 0.23\%       & 32                & 0.15\%         & 20.46            \\
    Convolution       & 3,113,888         & 91.76\%   & 187,003               & 56.37\%      & 3,152              & 14.10\%        & 59.32            \\
    Multiply          & 9,652            & 0.28\%    & 1,841                 & 0.64\%       & 489               & 2.19\%         & 3.76             \\
    Reshape           & 51,041           & 1.50\%    & 870                  & 0.30\%       & 103               & 0.46\%         & 8.39             \\
    Result            & 40              & 0.00\%    & 117                  & 0.04\%       & 0.5               & 0.00\%         & 207.42           \\
    Slice             & 30,338           & 0.89\%    & 576                  & 0.20\%       & 19                & 0.09\%         & 30.11            \\ \midrule
    Total             & 3,393,517         &           & 286,066               &              & 22,358             &                & 12.79            \\
    \bottomrule
  \end{tabular}

  \label{tab:32MM_075_96}

  \end{table*}

The most time-consuming operation is Convolution, representing almost
92\% of the execution, although it only accounts for 14\% of the
PMem accesses.
Convolution only accesses PMem every ${\sim}59$ DRAM accesses, which
indicates that DRAM caches the working set efficiently and therefore the PMem is not limiting the execution time of this operation.

The second most time-consuming operation is BoundedRelu, representing 3.7\% of the execution.
This operation is not supported by the encryption scheme and in a production scenario it would be performed by a remote trusted machine using the client-aided protocol discussed in Section~\ref{sec:hetransformer}.
With the current setup, the operation is performed on the same machine, which has to: decrypt, operate, and encrypt the result.
The encryption\slash decryption processes are strongly dependent on modified Fast Fourier Transform (FFT) operations, which feature non-sequential memory accesses.
This explains the high number of accesses to PMem in BoundedRelu when compared to the remaining operations.
To evaluate that this approach was not impacting the overall behavior of the application, we conducted an analysis of the operations immediately following the invocation of BoundedRelu.
We observe that the values of cache misses in these operations are similar to those of the functions of the same type prior to a BoundedRelu.
In a production scenario, we expect this value to be slightly lower because
not performing the operation would not invalidate as many cache
entries as we are currently observing.
However, there would be a higher latency in the unsupported operations
because of the need to communicate the data over the network.

We also found that Add and Multiply operations experience a ratio
between DRAM and PMem accesses below 5, which is notably low.
However, these only account for 1\% of the total time; hence, this low ratio does not significantly affect the overall execution.

\subsubsection{Analysis with Intel\textsuperscript{\textregistered}~VTune\textsuperscript{\texttrademark}~Platform Profiler}
\label{subsec:PlatformProfiler}

We used the
Intel\textsuperscript{\textregistered}~VTune\textsuperscript{\texttrademark}~Platform Profiler to
further confirm whether the workload used the underlying memory architecture efficiently.
The tool provides independent performance reports for each socket (among other components) but for
the sake of brevity we only show the profiling data from the socket that performs the 
application initialization.

\begin{figure*}\centering
  \subfloat[Optane cache hit ratio]{\includegraphics[width=0.9\textwidth]{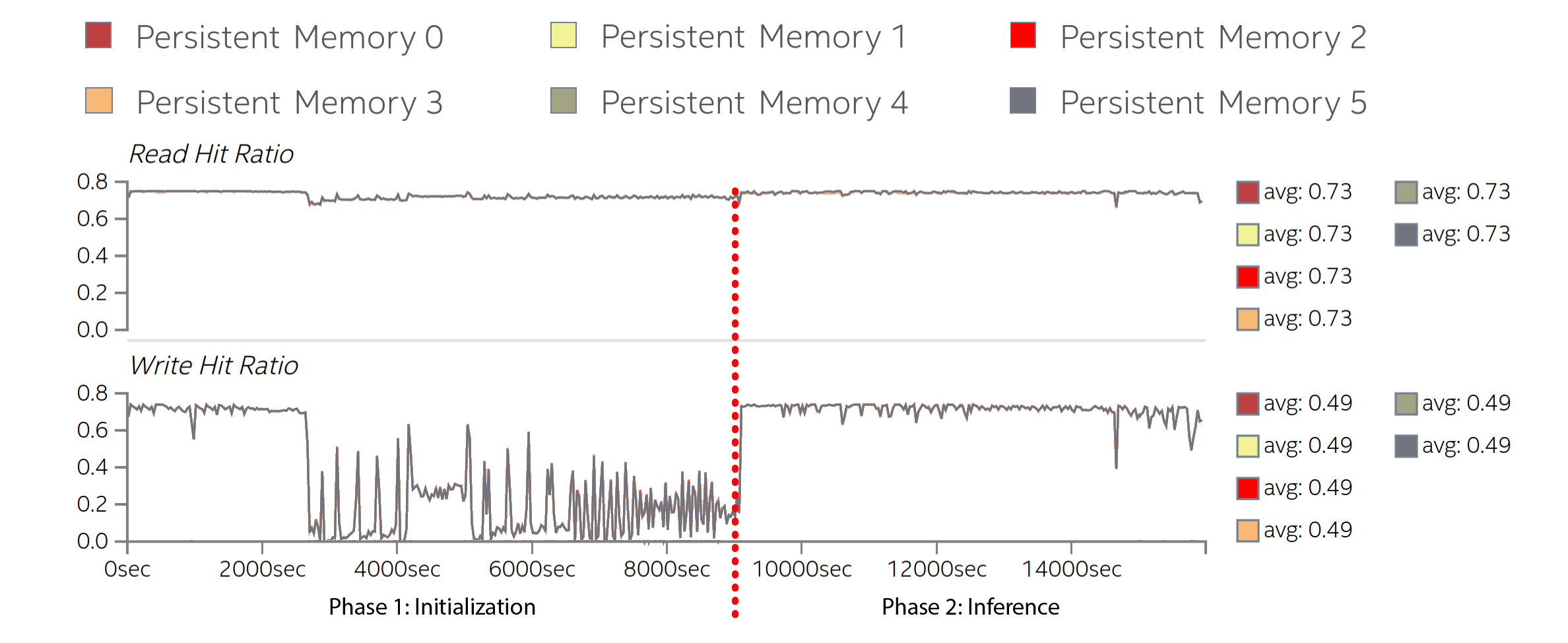}
    \label{fig:optanecache}}
\\
  \subfloat[DRAM traffic]{\includegraphics[width=0.9\textwidth]{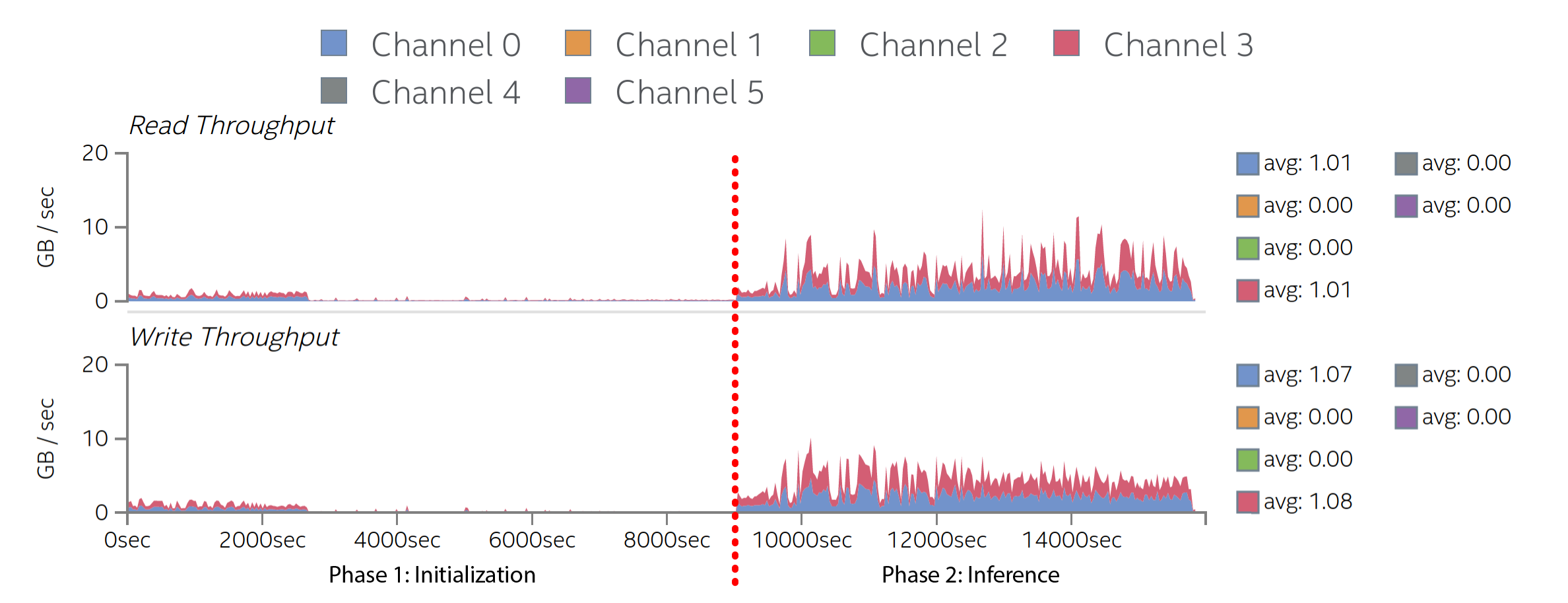}
    \label{fig:dram_traffic}}
\\
  \subfloat[PMem traffic]{\includegraphics[width=0.9\textwidth]{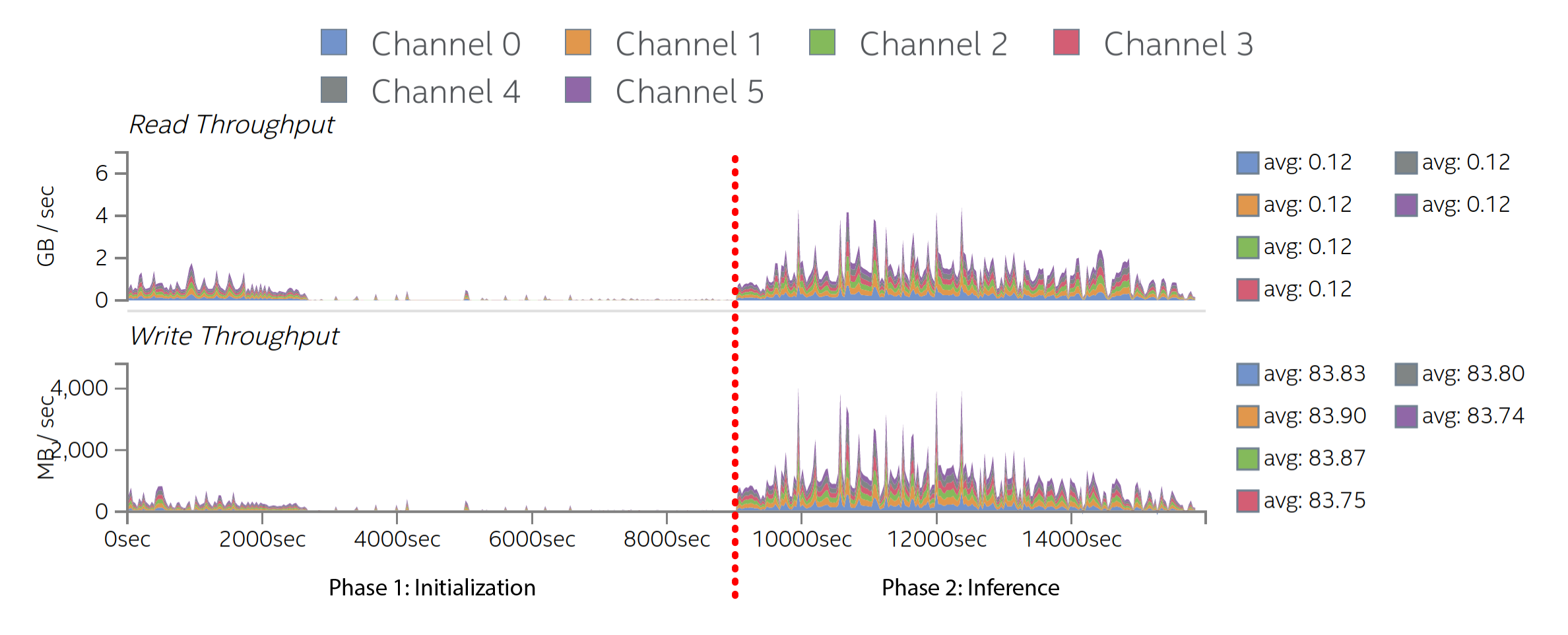}
    \label{fig:pmm_traffic}}
  \caption{Platform Profiler relevant data for MobileNetV2 (1.4, 224)
    in MM32.}
  \label{fig:platpro}
\end{figure*}

Figure~\ref{fig:platpro} depicts the metrics when the system runs the application
and the results evidence two execution phases.
The first phase corresponds to the initialization and optimization of the computation graph that describes the model.
This phase executes in a single core, which relates to the low overall activity observed during this phase.
The second phase corresponds to the inference, which in contrast to the initialization, runs mostly in parallel on all processor cores.
In the inference phase, and according to the profiling results, the
workload is fairly balanced among the sockets.
We present the results for the MobileNetV2 (1.4, 224) architecture, which are
highly similar to those of the smaller models.

Figure \ref{fig:optanecache} shows that the inference phase starts
a behavior corresponding to mostly sequential accesses.
As we described earlier, Intel Optane PMem is optimized for 256-byte
media operations and the PMem modules feature a media prefetch buffer that stores contiguous bytes.
In sequential access patterns, we expect a 0.75 hit ratio because the first data access causes a miss in the prefetch buffer but the three subsequent accesses hit on the buffer.
Again, the sequential access pattern exposed by the workload favors
the memory accesses and matches with the metrics reported by
Paraver. Sequential access patterns are justified in further detail in
Section~\ref{sec:appendix}.

To assess whether the persistent memory implementation poses a bandwidth
limitation, we have used the MM32 configuration.
In this situation, each DRAM DIMM features an individual bandwidth
\textit{circa} 21~GB/s and since there are two DIMMs per socket
populated, the bandwidth per socket is close to 42~GB/s (see Figure~\ref{fig:conf_mm32}).
Our profiling reveals that the average memory bandwidth in one socket is
2.02~GB/s for reading and and 2.15~GB/s for writing 
with peaks of around 20~GB/s (Figure~\ref{fig:dram_traffic}).
A similar behavior is observed in the sibling socket (but not shown in the figure).
Each PMem module, on the other hand, yields an individual bandwidth of 7.3~GB/s for reading and 2.4~GB/s for writing and since each socket is fully populated with six DIMMs, the bandwidth per socket is close to 43.8~GB/s for reading and 14.4~GB/s for writing.
Figure~\ref{fig:pmm_traffic} illustrates that neither writing nor reading bandwidth attain the maximum value.

These results are aligned with those exposed in
Section~\ref{subsec:Paraver} and reinforce our statement that
Intel Optane PMem yields efficient executions for
this type of workloads.

\subsection{ResNet-50}
\label{sec:res50results}

\begin{figure}\centering
  \begin{minipage}{.49\textwidth}\centering
    \includegraphics[width=1\textwidth]{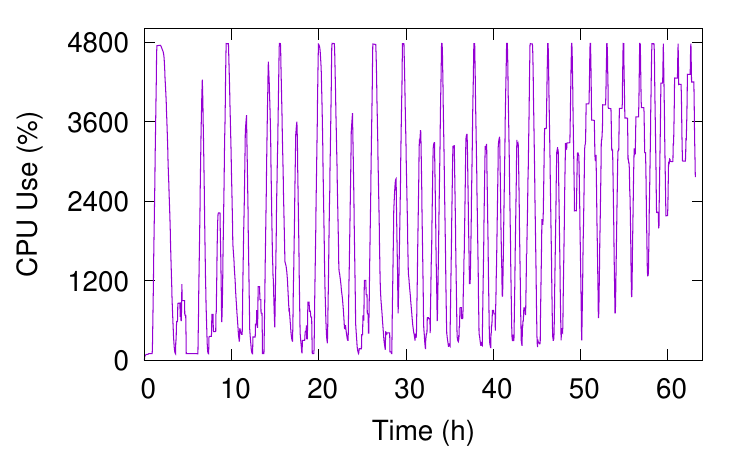}
    \caption{ResNet-50 CPU usage.}\label{fig:resnet_cpu}
  \end{minipage}
  \begin{minipage}{.49\textwidth}\centering
    \includegraphics[width=1\textwidth]{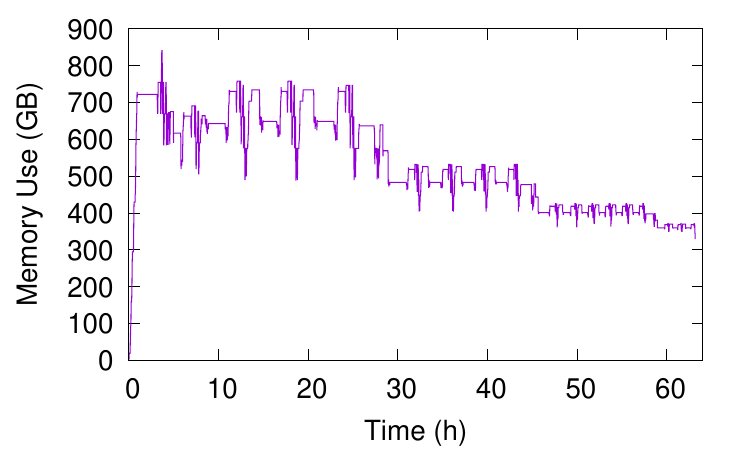}
    \caption{ResNet-50 memory use.}\label{fig:resnet_mem}
  \end{minipage}
\end{figure}

With our setup, we have been able to perform, for the first time,
inference using the ResNet-50 model, hence becoming the largest neural network
ever run using HE.
Leveraging a batch of 2,048 (the maximum value possible using the encryption parameters mentioned in Section~\ref{sec:hetransformer}) the complete execution lasts over 63 hours.
Figure~\ref{fig:resnet_cpu} shows the behavior of the CPU throughout the execution.
We observe the alternation of parallel functions (which use the 48
available cores) with sequential functions, following the expected fork-join model.
Since we are leveraging the use case of plaintext models (and encrypted
data), despite ResNet-50 being much more computationally demanding
than the considerably smaller MobileNetV2 (1.4, 224) model, the former requires a
slightly smaller amount of memory than the latter.
Figure~\ref{fig:resnet_mem} shows that the ResNet-50 peak memory consumption stands below 900~GB.

Due to the length of the execution, it has been impossible for us to
profile it with any of the the tools at our reach.
Since the predominant function in this type of networks is convolution,
and the access pattern for this function remains invariant with the input
size, we reasonably postulate that ResNet-50
leverages our persistent-memory-based memory setup as efficiently as
MobileNetV2. This postulate applies to any other
convolution-dominated model.

\subsection{Discussion: Access Sequentiality}
\label{sec:appendix}

Sections~\ref{sec:mobilenetresults} and~\ref{sec:res50results} show Intel Optane PMem yielding execution
efficiency for large HE models, due to cache-friendly sequential memory access patterns, particularly in the Convolution operation.
The Convolution operation is dominated by ciphertext--plaintext addition and ciphertext--plaintext multiplication, each of which features sequential memory accesses, implemented in SEAL~\cite{sealcrypto}.
Note that we do not consider the CKKS rescaling operation here, since its runtime impact is minimized by the use of lazy rescaling~\cite{boemer2019ngraph-he2}.
SEAL represents a ciphertext in RNS form as a $2 \times L \times N$-length vector of unsigned 64-bit integers.
However, ciphertext--plaintext operations require only one of the two ciphertext polynomials.
The plaintext argument in SEAL is represented as an $L \times N$-sized vector of unsigned 64-bit integers.
We use the optimization from nGraph-HE2~\cite{boemer2019ngraph-he2}, in the case where the plaintext encodes a single scalar, such that the plaintext argument is an $L$-length vector.

\subsubsection{Ciphertext--Plaintext Addition}
\label{sec:plain_addition}
CKKS ciphertext--plaintext addition in RNS form requires element-wise addition followed by modular reduction of two polynomials in which each sum is reduced with respect to the coefficient modulus $q_\ell$.

The following algorithm shows the pseudocode for ciphertext--plaintext scalar addition. The ciphertext argument is arranged in memory such that the two-dimensional memory accesses are sequential.

    \begin{algorithmic}[1]
        \Function{Add Cipher-Plain Scalar}{$c \in \mathbb{Z}^{L \times N}, p \in \mathbb{Z}^{L}, q \in \mathbb{Z}^L$}
        \For{ $\ell = 1$ to $L$}
        \State $tmp \gets p[\ell]$
        \For{ $n = 1$ to $N$}
        \State $c[\ell][n] \gets (c[\ell][n] +tmp) \text{\ mod\ } q[\ell]$
        \EndFor
        \EndFor
        \EndFunction
    \end{algorithmic}

\subsubsection{Ciphertext--Plaintext Multiplication}
\label{sec:plain_mult}
CKKS ciphertext--plaintext multiplication in RNS form requires
element-wise multiplication followed by modular reduction of two
polynomials in which each product is reduced with respect to the
coefficient modulus $q_\ell$. As in SEAL~\cite{sealcrypto},
nGraph-HE2~\cite{boemer2019ngraph-he2} uses Barrett reduction for
efficient modulus reduction. The following algorithm shows the pseudocode for ciphertext--plaintext scalar multiplication, in the case where the coefficient modulus is less than 32 bits, as in our setting. The ciphertext argument is arranged in memory such that the two-dimensional memory accesses are sequential.

    \begin{algorithmic}[1]
        \Function{Multiply Cipher-Plain 32-bit}{$c \in \mathbb{Z}^{L \times N}, p \in \mathbb{Z}^{L}, q \in \mathbb{Z}^L$, $r \in \mathbb{Z}^L$}
        \For{ $\ell = 1$ to	$L$}
        \State $tmp \gets p[\ell]$
        \For{ $n = 1$ to $N$}
        \State $uint64\ z \gets c[\ell][n]* tmp$ 
        \State $c[\ell][n] \gets BarrettReduction64(z, q[\ell], r[\ell])$
        \EndFor
        \EndFor
        \EndFunction
    \end{algorithmic}

These algorithms show that the
ciphertext accesses are sequential in memory. This yields the high
ratio of cache hits seen in Figure~\ref{fig:optanecache} and supports
the high efficiency experienced using Intel Optane PMem.

\section{Conclusions}
\label{sec:conclusions}

In this article we report running inference for the largest DNN models to
date leveraging HE. For the first time in the literature, we use
recently-emerged persistent memory technology as an enabler for such large memory
footprints.  Our novel analysis reveals that DNN inference leveraging HE
features memory access patterns that yield efficient use of the Intel
Optane PMem in Memory Mode.  Sequential data accesses in the
most common operations enable the accessed data to be efficiently
cached in the DRAM, but also to make efficient use of the hardware prefetch
buffers in the PMem. We propose a memory configuration with reduced
DRAM size that is cost-efficient, equipping 1/3 of the DRAM bandwidth
while reducing merely 10\% performance with respect to a
fully-populated system.

\ifCLASSOPTIONcompsoc
  \section*{Acknowledgments}
\else
  \section*{Acknowledgment}
\fi

We would like to thank Jesus Labarta from BSC and Steve Scargall from Intel for their insightful and productive comments.

\ifCLASSOPTIONcaptionsoff
  \newpage
\fi



\bibliographystyle{IEEEtran}
\bibliography{references}
%



%

\begin{IEEEbiography}[{\includegraphics[width=1in,clip,keepaspectratio]{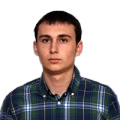}}]{Guillermo Lloret-Talavera}
is a  Jr. Research  Engineer  at  the  Barcelona  Supercomputing  Center  (BSC),  working  in  the Accelerators and Communications for HPC team. His interests include performance tuning for heterogeneous memory systems and deep learning frameworks.
\end{IEEEbiography}

\begin{IEEEbiography}[{\includegraphics[width=1in,clip,keepaspectratio]{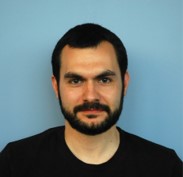}}]{Marc
  Jorda}
is a Research Engineer at the Barcelona Supercomputing Center (BSC), working in the Accelerators and Communications for HPC team. His interests include performance tuning for heterogeneous memory systems, GPU-enabled applications, and deep learning frameworks.
\end{IEEEbiography}

\begin{IEEEbiography}[{\includegraphics[width=1in,clip,keepaspectratio]{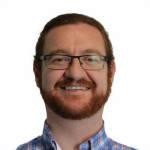}}]{Dr.
  Harald Servat}
is an HPC system enthusiast with strong knowledge in monitoring systems, parallel programming models, compilers and computer architecture. He currently works at Intel Corp. on code modernization topics for the next generation HPC systems. Before that, he was the maintainer of the instrumentation library for the BSC performance tools suite (Extrae) while adapting it to new technologies and pursuing large scalability. In 2015, he received his Ph.D. in providing instantaneous metrics combining coarse-grain instrumentation and sampling techniques. During his research, he explored the performance of several in-production applications and applied code transformations to increase the application performance.
\end{IEEEbiography}

\begin{IEEEbiography}[{\includegraphics[width=1in,clip,keepaspectratio]{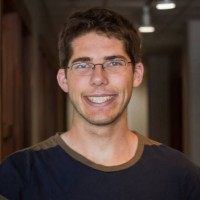}}]{Fabian
    Boemer}
is a research scientist at Intel Corporation. He received his Master's degree from Stanford University in Computational and Mathematical Engineering in 2018. Fabian's interests lie in privacy-preserving machine learning, in particular homomorphic encryption (HE). Fabian maintains the nGraph-HE software library (ngra.ph/he), which enables deep learning on homomorphically encrypted data.
\end{IEEEbiography}

\begin{IEEEbiography}[{\includegraphics[width=1in,clip,keepaspectratio]{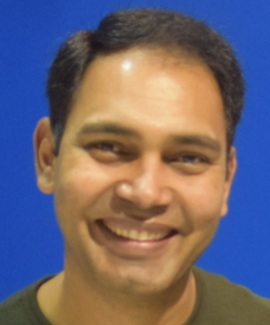}}]{Chetan
  Chauhan}
is Optane Component Architect in Intel's Nonvolatile Memory Storage
Group. His focus is on developing\slash simulating power and performance models for pathfinding new system architecture for Optane. He is very interested in exploring how Optane can help AI use cases like homomorphic encryption. He has received his Master's in Computer Engineering from Syracuse University.
\end{IEEEbiography}

\begin{IEEEbiography}[{\includegraphics[width=1in,clip,keepaspectratio]{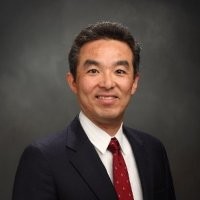}}]{Dr.
  Shigeki Tomishima}
received his B.S. and M.S. degrees in Solid State Physics and his
Ph.D. degree in Electric Engineering from Osaka University, Osaka,
Japan in 1988, 1990 and 2002, respectively. After 20 years of DRAM memory
array and architecture research in DRAM industry, he joined Intel
Corporation to work on the embedded DRAM project. In 2014, he
joined Intel Labs\slash Memory Architecture Lab to start working on advanced
memory architecture research for future computing systems, the
emerging memory technology including Compute-Near-Memory and
Compute-In-Memory concepts. He has authored and coauthored more than 15
international conference papers, 10 journal papers, 15 invited
talks, and holds 129 issued U.S. patents. He received Best Paper
Reviewer Award from IEEE CAS in 2016 and serves as TPC member on A-SSCC, VLSI-DAT, ISQED, and as a paper reviewer on JSSC, SSC-L, TVLSI, MICRO, ISCA, AICAS, and ISCAS.
\end{IEEEbiography}

\begin{IEEEbiography}[{\includegraphics[width=1in,clip,keepaspectratio]{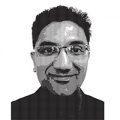}}]{Nilesh
  N. Shah}
directs  Computational Storage at Intel's Nonvolatile Memory Storage
Group's Data Center division and likes hacking privacy preserving machine learning code for fun. He is specially interested in homomorphic encryption, federated deep learning and mixing and matching storage technologies with heterogeneous accelerators.
\end{IEEEbiography}

\begin{IEEEbiography}[{\includegraphics[width=1in,clip,keepaspectratio]{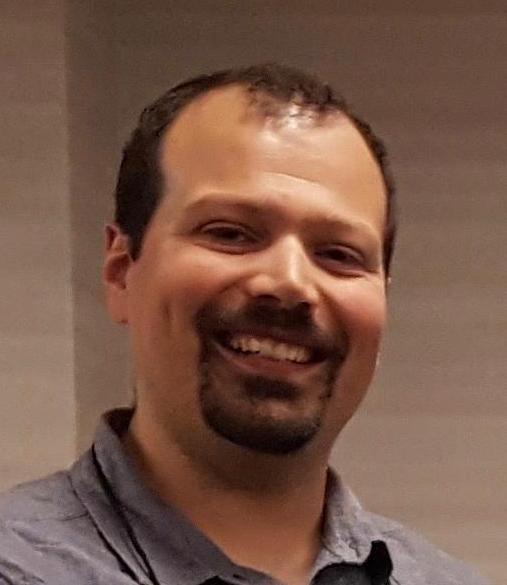}}]{Dr.
    Antonio J. Pe\~na}
holds a BS+MS degree in Computer Engineering (2006), and MS and PhD
degrees in Advanced Computer Systems (2010, 2013), from Universitat
Jaume I de Castell\'o, Spain. He is currently a Sr. Researcher at the Barcelona Supercomputing Center (BSC), where he leads the ``Accelerators and Communications for HPC'' team. Antonio is a former Marie Sklodowska-Curie Fellow and Juan de la Cierva Fellow. He is a recipient of the 2017 IEEE TCHPC Award for Excellence for Early Career Researchers in HPC. His research interests in the area of runtime systems and programming models for HPC include resource heterogeneity and communications.
\end{IEEEbiography}



\vfill


\end{document}